\documentclass[12pt]{article}
\usepackage{epsf,amsfonts,amssymb,epsfig,amsmath}
\renewcommand{\text}[1]{#1}

\newcommand{\be}{\begin{equation}}
\newcommand{\ee}{\end{equation}}
\newcommand{\ben}{\begin{displaymath}}
\newcommand{\een}{\end{displaymath}}
\newcommand{\bea}{\begin{eqnarray}}
\newcommand{\eea}{\end{eqnarray}}
\newcommand{\bean}{\begin{eqnarray*}}
\newcommand{\eean}{\end{eqnarray*}}
\newcommand{\nn}{\nonumber \\}
\newcommand{\ba}{\begin{array}}
\newcommand{\ea}{\end{array}}
\newcommand{\bi}{\begin{itemize}}
\newcommand{\ei}{\end{itemize}}

\newcommand{\reef}[1]{(\ref{#1})}

\def\e{\epsilon}

\def\e{\epsilon}

\newcommand{\bbR}{{\mathbb{R}}}

\newcommand{\sla}{\slash\!\!\!\!}
\begin{document}

\makeatletter
\renewcommand{\theequation}{\thesection.\arabic{equation}}
\@addtoreset{equation}{section}
\makeatother

\baselineskip 18pt

\begin{titlepage}

\vfill

\begin{flushright}
Imperial/TP/2009/JG/02\\
DESY 09-006\\
\end{flushright}

\vfill

\begin{center}
   \baselineskip=16pt
   {\Large\bf  Supersymmetric solutions for\\ non-relativistic holography}
  \vskip 1.5cm
      Aristomenis Donos$^1$ and Jerome P. Gauntlett$^2$\\
   \vskip .6cm
      \begin{small}
      $^1$\textit{DESY Theory Group, DESY Hamburg\\
        Notkestrasse 85, D 22603 Hamburg, Germany}
        %E-mail: j.gauntlett, d.waldram@imperial.ac.uk}
        \end{small}\\*[.6cm]
        \begin{small}
      $^2$\textit{Theoretical Physics Group, Blackett Laboratory, \\
        Imperial College, London SW7 2AZ, U.K.}
        %E-mail: j.gauntlett, d.waldram@imperial.ac.uk}
        \end{small}\\*[.6cm]
      \begin{small}
      $^2$\textit{The Institute for Mathematical Sciences, \\
        Imperial College, London SW7 2PE, U.K.}
        %E-mail: j.gauntlett, d.waldram@imperial.ac.uk}
        \end{small}\\*[.6cm]
   \end{center}

\vfill

\begin{center}
\textbf{Abstract}
\end{center}

\begin{quote}
We construct families of supersymmetric solutions of type IIB and $D=11$ supergravity that
are invariant under the non-relativistic conformal algebra for various values of dynamical
exponent $z\ge 4$ and $z\ge 3$, respectively. The solutions
are based on five- and seven-dimensional Sasaki-Einstein manifolds
and generalise the known solutions with dynamical exponent $z=4$ for the type IIB case and $z=3$
for the $D=11$ case, respectively.
\end{quote}

\vfill

\end{titlepage}
\setcounter{equation}{0}

%%%%%%%%%%%%%%%%%%%%%%%%%%%%%%%%%%%%%%%%%%%%%%%%%%%%%%%%%%%%%%%%%%%%%%%
%\tableofcontents
%%%%%%%%%%%%%%%%%%%%%%%%%%%%%

\section{Introduction}

There has recently been much interest in finding holographic realisations of
systems invariant under the non-relativistic conformal algebra starting with the work
\cite{Son:2008ye}, \cite{Balasubramanian:2008dm} and discussed further in related work \cite{Sakaguchi:2008rx}-\cite{pal}.
Such systems are invariant under Galilean transformations, generated by time and spatial translations, spatial
rotations, Galilean boosts and a mass operator, which is a central element of the algebra,
combined with scale transformations.
If $x^+$ is the time coordinate, and ${\bf x}$ denotes $d$ spatial coordinates,
the scaling symmetry acts as
\be\label{scalsym}
{\bf x}\to \mu {\bf x},\qquad x^+\to \mu^z x^+\ ,
\ee
where $z$ is called the dynamical exponent. When $z=2$ this non-relativistic conformal
symmetry can be enlarged to an invariance under the Schr\"odinger algebra which includes an additional
special conformal generator.

The solutions found in \cite{Son:2008ye}, \cite{Balasubramanian:2008dm} with $d=2$ and $z=2$ were subsequently
embedded into type IIB string theory in \cite{Herzog:2008wg},\cite{Maldacena:2008wh},\cite{Adams:2008wt}
and were based on an arbitrary five-dimensional Sasaki-Einstein manifold, $SE_5$.
The work of \cite{Maldacena:2008wh} also constructed type IIB solutions with $d=2$ and $z=4$
and again these were constructed using an arbitrary $SE_5$. It was also shown in \cite{Maldacena:2008wh}
that the solutions with $z=2$ and $z=4$ can be obtained from a five dimensional theory with a massive vector field
after a Kaluza-Klein reduction on the $SE_5$ space \cite{Maldacena:2008wh}. This procedure was generalised to solutions of
$D=11$ supergravity in \cite{gkvw}: using a similar KK reduction on an arbitrary seven-dimensional Sasaki-Einstein space, $SE_7$,
solutions with non relativistic conformal symmetry with $d=1$ and $z=3$ were found.

The type IIB solution of
\cite{Herzog:2008wg},\cite{Maldacena:2008wh},\cite{Adams:2008wt}
with $z=2$ do not preserve any supersymmetry
\cite{Maldacena:2008wh}. One aim of this note is to show that, by
contrast, the type IIB solutions of \cite{Maldacena:2008wh} with
$z=4$ and the $D=11$ solutions of \cite{gkvw} with $z=3$ are both
supersymmetric and generically preserve two supersymmetries. A
second aim is to generalise both of these supersymmetric solutions
to different values of $z$. We will construct new
supersymmetric solutions using eigenmodes of the Laplacian acting on
one-forms on the $SE_5$ or $SE_7$ space. If the eiegenvalue is $\mu$
then we obtain type IIB solutions with $z=1+\sqrt{1+\mu}$ and $D=11$
solutions with $z=1+\tfrac{1}{2}\sqrt{4+\mu}$. This gives
rise to type IIB solutions with $z\ge 4$ and $D=11$ solutions with $z\ge 3$, respectively.
For the case of $S^5$ we get solutions with $z=4,5,\dots $ while for the case of $S^7$ we get solutions with $z=3,3\tfrac{1}{2},4,\dots$
and both of these preserve 8 supersymmetries.

Our constructions have some similarities with the construction of type IIB solutions
in \cite{Hartnoll:2008rs} that were based on eigenmodes of the Laplacian acting on scalar functions on the $SE_5$ space.
Our IIB solutions preserve the same supersymmetry and we show how our solutions can be superposed with those
of \cite{Hartnoll:2008rs} while maintaining a scaling symmetry. An analogous superposition is possible for
the $D=11$ solutions, which we shall also describe.

\section{The type IIB solutions}

The ansatz for the type IIB solutions we shall consider is given by\\
\bea\label{iibsol}
ds^2&=&\frac{dr^2}{r^2}+r^2\left[2dx^+dx^-+dx_1^2+dx_2^2\right]+ds^2(SE_5)+2r^2C
dx^+\nn
F_5&=&4r^3dx^+\wedge dx^-\wedge dr\wedge dx_1\wedge dx_2+4Vol(SE_5)\nn
&-&dx^+\wedge\left[*_{CY_3} dC+d(r^4C)\wedge dx_1\wedge dx_2\right]
\eea
where $SE_5$ is an arbitrary five-dimensional Sasaki-Einstein space and
the metric $ds^2(SE_5)$ is normalised so that the Ricci tensor is equal to four
times the metric (i.e. the same normalisation as that of a unit radius five-sphere).
Recall that the metric cone over the $SE_5$,
\be
ds^2(CY_3)=dr^2+r^2ds^2(SE_5)\ ,
\ee
is Calabi-Yau. The K\"ahler form
%and the holomorphic $(3,0)$ form
on the $CY_3$ is
denoted $\omega_{ij}$ and
%$\Omega_{ijk}$, respectively. The
the complex structure is defined\footnote{While this is standard in the physics literature, often in the maths literature
$J_i{}^j=-\omega_{ik}g^{kj}$.}
 by $J_i{}^j=\omega_{ik}g^{kj}$, where $g_{ij}$ is the Calabi-Yau cone metric. We will define the one-form
$\eta$, which is dual to the Reeb vector on $SE_5$ by
\be
\eta_i=-J_i{}^j\left(d \log r \right)_j\ .
\ee

The one-form $C$ is a one-form on the $CY_3$ cone.
When $C=0$ we have the standard $AdS_5\times SE_5$ solution
of type IIB which, in general, preserves eight supersymmetries (four Poincar\'e and four superconformal), corresponding
to an $N=1$ SCFT in $d=4$. More generally, we can deform this solution by
choosing $C\ne 0$ provided that $dC$ is co-closed on $CY_3$:
\be
d*_{CY}dC=0\ .
\ee
With this condition, $F_5$ is closed and in fact it is also
sufficient for the type IIB Einstein equations to be satisfied.
As we will show these solutions preserve one half of the Poincar\'e supersymmetries.
Note that the solution is invariant under the transformation
\be\label{gt}
x^-\to x^--\Lambda,\qquad C\to C+d\Lambda
\ee
for some function $\Lambda$ on the CY cone. Thus, if $dC=0$, we can remove $C$, at least locally, by
such a transformation.

We will look for solutions where the one-form $C$ has weight $\lambda$ under the action of $r\partial_r$.
Then it is straightforward to check, following \cite{Son:2008ye} and \cite{Balasubramanian:2008dm}
that our solution is invariant under non-relativistic conformal transformations with two spatial dimensions
$x^1$, $x^2$ and dynamical exponent $z=2+\lambda$.
For example the scaling symmetry is acting as in \reef{scalsym} combined with $r\to \mu^{-1}r$, $x^-\to\mu^{2-z}x^-$.
Following the analysis 
of closed and co-closed two forms on cones
(such as $dC$) in appendix A of \cite{Martelli:2008cm} 
we consider solutions constructed from a co-closed one-form $\beta$ on the $SE_5$ space that is an eigenmode of the Laplacian
$\Delta_{SE}=(d^\dagger d+d d^\dagger)_{SE}$:
\be\label{newcee}
C=r^\lambda \beta,\qquad \Delta_{SE}\beta=\mu\beta,\qquad d^\dagger\beta=0\ .
\ee
It is straightforward to check that $dC$ is co-closed providing that
$\mu=\lambda(\lambda+2)$. For our applications we choose the branch
$\lambda=-1+\sqrt{1+\mu}$ leading to solutions with
\be
z=1+\sqrt{1+\mu}\ .
\ee
A general result valid for any five-dimensional Einstein space, normalised as we have, is that for co-closed 1-forms $\mu\ge 8$
and $\mu=8$ holds iff the 1-form is dual to a Killing vector (see section 4.3 of \cite{Duff:1986hr}). Thus in general our construction leads to solutions
with
\be\label{bd}
z\ge 4\ .
\ee
Since all $SE_5$ manifolds have at least the Reeb Killing vector, dual to the one-form $\eta$, this bound is
always saturated. Indeed the solution of \cite{Maldacena:2008wh} with $z=4$ is in our class. Specifically it can be
obtained by setting $C=\sigma r^2\eta$ (and redefining $x^-\to -x^-/2$): one can explicitly check
that $\eta$ is co-closed on $SE_5$ and is an eigenmode of  $\Delta_{SE}$ with eigenvalue $\mu=8$.
Note that for this solution the two-form $dC$ is proportional to the K\"ahler-form of the Calabi-Yau cone:
$dC=2\sigma \omega$.

On  $S^5$ the spectrum of $\Delta_{S^5}$ acting on one-forms is well known and we have $\mu=(s+1)(s+3)$ for $s=1,2,3\dots$ (see for example \cite{Kim:1985ez} eq (2.20))
leading to $\lambda=s+1$ and hence new classes of solutions with $z=4,5,6\dots$. Note that these solutions come in families, transforming in the
$SO(6)$ irreps ${\bf 15}$, ${\bf 64}$, ${\bf 175}$, $\dots$.
To obtain similar results for $T^{1,1}$ one can consult \cite{Ceresole:1999ht}.

We now discuss a construction that can be used when the spectrum of the Laplacian acting on functions is known, but
not acting on one-forms. For example, the scalar Laplacian was studied in \cite{Kihara:2005nt} for the $Y^{p,q}$ metrics \cite{Gauntlett:2004yd},
but as far as we know it has not been discussed acting on one-forms.
Specifically we construct $(1,1)$ forms $dC$ on the CY cone using scalar functions $\Phi$ on the cone as follows. We write
\be\label{expc}
C_i=J_i{}^j\partial_j\Phi
\ee
for some
function $\Phi$ on $CY_3$. A short calculation shows that
if
\be\label{cond}
\nabla^2_{CY}\Phi=\alpha
\ee
for some constant $\alpha$ then $dC$ is co-closed. The two-form $dC$ is a $(1,1)$ form on $CY_3$ and it
is primitive, $J^{ij}dC_{ij}=0$, if and only if $\alpha=0$.
Observe that the solution of \cite{Maldacena:2008wh} with $z=4$ fits into this class by taking
$\Phi=-\sigma r^2/2$ and $\alpha=-6\sigma$, leading to $C=\sigma r^2\eta$.

We now consider solutions with $\alpha=0$, corresponding to harmonic functions\footnote{Note that in general the
one-form $C$ defined in \reef{expc} has a component in the $dr$ direction, unlike in \reef{newcee}. However, locally we can remove it by
a transformation of the form \reef{gt}. Also, one can directly show that the resulting one-form $\beta$ is co-closed
on the $SE_5$ space.}
on the CY cone with $dC$ $(1,1)$ and primitive.
We next write
\be\label{expphi}
\Phi=r^\lambda f
\ee
where $f$ is a function on the $SE_5$ space satisfying
\be
-\nabla^2_{SE_5}f=kf
\ee
with $k=\lambda(\lambda+4)$ (see e.g. \cite{Gauntlett:2006vf}).
For the solutions of interest we choose the branch $\lambda=-2+\sqrt{4+k}$
leading to $z=\sqrt{4+k}$. For the special case of the five-sphere
we can check with the results that we obtained above. 
The eigenfunctions $f$ on the five-sphere are given by spherical harmonics with
$k=l(l+4)$, $l=1,2,\dots$ and hence $z=l+2$. The $l=1$ harmonic appears to violate the bound \reef{bd}.
However, it is straightforward to see that the construction for $l=1$ leads to $dC=0$ for which $C$ can be removed by a transformation
of the form \reef{gt}.
Thus for $S^5$ we should consider $l\ge 2$ leading to solutions with $z=4,5,\dots$, as above.
It is worth pointing out that for higher values of $l$ some of the eigenfunctions will also
lead to closed $C$: if we consider the harmonic function on $\bbR^6$ given by $x^{i_1}\dots x^{i_l}c_{i_1\dots i_l}$ where $c$ is symmetric and traceless
then, with $J=dx^1\wedge dx^2+dx^3\wedge dx^4+dx^5\wedge dx^6$ we see that $dC=0$ if
$J_{[i}{}^jc_{k]ji_3\dots i_l}=0$.

\subsection{Supersymmetry}

We introduce the frame
\bea e^+&=&rdx^+\nn
e^-&=&r(dx^-+C)\nn e^2&=&rdx_1\nn e^3&=&rdx_2\nn
e^4&=&\frac{dr}{r}\nn e^m&=&e^m_{SE},\qquad m=5,\dots,9 \eea
where $e^m_{SE}$ is an orthonormal frame for the $SE_5$ space.
We can write
\begin{align}
F_{5}=& B_{5}+\ast_{10}B_{5}\\
B_{5}=&4 e^{+}\wedge e^{-}\wedge e^{2}\wedge e^{3}\wedge e^{4}-{r}e^{+}\wedge dC\wedge e^{2}\wedge e^{3}
\end{align}
where we have chosen $\epsilon_{+-23456789}=+1$. The Killing spinor equation can be written
\bea
D_M\epsilon+\frac{i}{16}\sla{F}\Gamma_M\epsilon=
D_M\epsilon+\frac{i}{2}\sla{B}\Gamma_M\epsilon=0\ .
\eea
We are using the conventions for type IIB supergravity \cite{Schwarz:1983qr}\cite{Howe:1983sra}
as in \cite{Gauntlett:2005ww} and
in particular, $\Gamma_{11}=\Gamma_{+-23456789}$ with the chiral IIB spinors satisfying
$\Gamma_{11}\epsilon=-\epsilon$.
%Spin connection
%\begin{align}
%\omega_{-4}&=e^{+}\\
%\omega_{+4}&=e^{-}+\frac{1}{2}r^{2}dC_{4\mu}dx^{\mu}+\frac{1}{2}rdh_{4} e^{+}\\
%\omega_{+m}&=\frac{1}{2}r^{2}dC_{m\mu}dx^{\mu}+\frac{1}{2}rdh_{m} e^{+}\\
%\omega_{24}&=e^{2}\\
%\omega_{34}&=e^{3}\\
%\omega_{mn}&=\omega_{\left(SE\right)mn}-\frac{1}{2}rdC_{mn}e^{+}\\
%\omega_{m4}&=-\frac{1}{2}rdC_{m4}e^{+}
%\end{align}

If $\epsilon$ are the Killing spinors for the $AdS_5\times SE_5$ solution,
then we find that we must also impose that
\bea
\Gamma^{+-23}\epsilon &=&i\epsilon\nn
\Gamma^{+}\epsilon &=&0 \label{IIB_proj1}\ .
\eea
The first condition maintains the Poincar\'e supersymmetries but breaks all
of the superconformal supersymmetries (this can be explicitly checked using, for example,
the results of \cite{Lu:1998nu}). The second condition breaks a further half of these\footnote{That we preserve
the Poincar\'e supersymmetries suggests that we can extend our solutions away from the near horizon limit of the D3-branes.
This is indeed the case but we won't expand upon that here.}.
Thus when $dC\ne 0$, we preserve two Poincar\'e supersymmetries for a generic $SE_5$ and this
is increased to eight Poincar\'e supersymmetries for $S^5$.

\section{The $D=11$ solutions}
The construction of the $D=11$ solutions is very similar.
We consider the ansatz for D=11 supergravity solutions:
\bea\label{11solr}
ds^2&=&\frac{d\rho^2}{4\rho^2}+\rho^2\left[2dx^+dx^-+dx^2\right]+ds^2(SE_7)+2\rho^2C
dx^+\nn G&=&-3\rho^2dx^+\wedge dx^-\wedge d\rho\wedge dx+dx^+\wedge dx\wedge d(\rho^3C)
\eea where
$SE_7$ is a seven-dimensional Sasaki-Einstein space and $ds^2(SE_7)$ is normalised
so that the Ricci tensor is equal to six times the metric (this is the normalisation of
a unit radius seven-sphere).
It is convenient to change coordinates via $\rho=r^2$ to
bring the solution to the form \bea\label{11sol}
ds^2&=&\frac{dr^2}{r^2}+r^4\left[2dx^+dx^-
+dx^2\right]+ds^2(SE_7)+2r^4C dx^+\nn G&=&-6r^5dx^+\wedge dx^-\wedge dr
\wedge dx+dx^+\wedge dx\wedge d(r^6C)\ .
\eea In these coordinates the cone metric \be
ds^2_{CY}=dr^2+r^2 ds^2(SE_7) \ee is a metric on Calabi-Yau
four-fold. We will use the same notation for the $CY$ space
as in the previous section.

When the one-form $C$ is zero we have the standard
$AdS_4\times SE_7$ solution of $D=11$ supergravity that, in general, preserves
eight supersymmetries. We again find that all the equations of
motion are solved if $C$ is a one-form on $CY_4$ and the two-form
$dC$ is co-closed \be d*_{CY}dC=0\ .\ee
The solutions are again invariant under the transformation \reef{gt}.
We will consider solutions where the one-form $C$ has weight
$\lambda$ under the action of $r\partial_r$, corresponding to
dynamical exponent $z=2+\lambda/2$.
As before, using the results in
appendix A of \cite{Martelli:2008cm}, we consider solutions
constructed from a co-closed one-form $\beta$ on the $SE_7$ space
that is an eigenmode of the Laplacian $\Delta_{SE}$:
\be\label{newcee11}
C=r^\lambda \beta,\qquad \Delta_{SE}\beta=\mu\beta, \qquad d^\dagger\beta=0\ .
\ee
One can check that $dC$ is co-closed providing that
$\mu=\lambda(\lambda+4)$. For our applications we choose the branch
$\lambda=-2+\sqrt{4+\mu}$ leading to solutions with
\be z=1+\tfrac{1}{2}\sqrt{4+\mu}\ .\ee
A
general result valid for any seven-dimensional Einstein space, normalised as we have,
is that for co-closed 1-forms $\mu\ge 12$ and $\mu=12$ holds iff the
1-form is dual to a Killing vector (see section 4.3 of \cite{Duff:1986hr}). Thus in
general our construction leads to solutions with \be\label{bd11}
z\ge 3 \ee and the bound is again saturated for all $SE_7$ spaces.
Observe that the solutions of \cite{gkvw} with $z=3$ fit into this
class. Specifically they are obtained by setting $C=\sigma
r^2\eta$ (after redefining $x\to x/2$ and $x^-\to -x^-/8$). On
$S^7$ the spectrum of $\Delta_{S^7}$ is well known and we have
$\mu=s(s+6)+5$ for $s=1,2,3\dots$ (see for example \cite{Duff:1986hr}
eq (7.2.5)) leading to $\lambda=1+s$ and hence new
classes of solutions with $z=3,3\tfrac{1}{2},4,\dots$. These solutions come in families transforming in
the $SO8)$ irreps ${\bf 28}$, ${\bf 160_v}$, ${\bf 567_v}$, $\dots$ .
Results on the spectrum
of the Laplacian on some homogeneous $SE_7$ spaces can be found in
\cite{Fabbri:1999mk},\cite{Merlatti:2000ed},\cite{Termonia:1999cs}.

As before we can construct $(1,1)$ co-closed two-forms $dC$ using scalar functions $\Phi$ on $CY_4$
We write
\be\label{expc11}
C_i=J_i{}^j\partial_j\Phi,\qquad \nabla^2_{CY}\Phi=\alpha~\ .
\ee
and $dC$ is again primitive if and only if $\alpha=0$. The
solutions of \cite{gkvw} with $z=3$ arise by taking $\Phi=\sigma r^2$ and $\alpha=-8\sigma$
leading to $C=\sigma r^2\eta$.
We now focus on solutions with $\alpha=0$, corresponding to harmonic functions on the CY cone.
We take
\be\label{expphi11}
\Phi=r^\lambda f
\ee
where $f$ is a function on the $SE_7$ space satisfying
\be
-\nabla^2_{SE_7}f=kf
\ee
with $k=\lambda(\lambda+6)$. For our applications we choose the branch
$\lambda=-3+\sqrt{9+k}$ leading to solutions with
$z=\tfrac{1}{2}+\tfrac{1}{2}\sqrt{9+k}$.
For example, on the seven-sphere the eigenfunctions $f$ are given by spherical harmonics with
$k=l(l+6)$ with $l=1,2,\dots$ and hence $z=2+l/2$.
Excluding the $l=1$ harmonic, as it can be removed by a transformation of the form \reef{gt},
for $S^7$ we are left with solutions with $z=3,7/2,4,\dots$, as above.

\subsection{Supersymmetry}

We introduce a frame \bea e^+&=&r^2dx^+\nn
e^-&=&r^2(dx^-+C)\nn e^2&=&r^2dx\nn
e^3&=&\frac{dr}{r}\nn e^m&=&e^m_{SE},\qquad m=4,\dots,10\ . \eea
We thus have
\bea G&=&6 e^+\wedge e^-\wedge e^2\wedge e^3+r^2 e^+\wedge e^2\wedge dC\nn
\ast_{11}G&=&-6Vol(SE_7)+dx^+ *_{CY}dC \eea
where we have chosen the orientation $\epsilon_{+-23....10}=+1$.

The Killing spinor equation can be written as
\begin{equation}\label{killing}
\nabla_M\e+\frac{1}{288}[\Gamma{_M}{^{N_1N_2N_3N_4}}
-8\delta{_M^{N_1}}\Gamma^{N_2N_3N_4}]G_{N_1N_2N_3N_4}\e=0\ .
\end{equation}
We are using the conventions for $D=11$ supergravity \cite{Cremmer:1978km}
as in \cite{gp}
and in particular $\Gamma_{+-2345678910}=+1$.

If $\epsilon$ are the Killing spinors arising for the $AdS_4\times SE_7$ solution,
then we find that we must also impose that
\bea
\Gamma^{+-2}\epsilon &=&-\epsilon\nn
\Gamma^{+}\epsilon &=&0\ . \label{11d_proj}
\eea
The first condition maintains the Poincar\'e supersymmetries but breaks all
of the superconformal supersymmetries. The second condition breaks a further half of these.
Thus when $dC\ne 0$, we preserve two Poincar\'e supersymmetries for a generic $SE_7$ and this
is increased to eight Poincar\'e supersymmetries for $S^7$.

\subsection{Skew-Whiffed Solutions}
If $AdS_4\times SE_7$ is a supersymmetric solution of $D=11$ supergravity,
then if we ``skew-whiff'' by reversing the sign of the flux (or equivalently changing the orientation of $SE_7$)
then apart from the special
case when the $SE_7$ space is the round $S^7$, all supersymmetry is broken \cite{Duff:1984sv}.
Despite the lack of supersymmetry, such solutions are known to be perturbatively
stable \cite{Duff:1984sv}. 
Similarly, if we reverse the sign of the flux in our new solutions \reef{11sol}, we will
obtain solutions of $D=11$ supergravity that will generically not preserve any supersymmetry.

\section{Further Generalisation}
We now discuss a further generalisation of the solutions that we have
considered so far,
preserving the same amount of supersymmetry, which incorporate the construction
of \cite{Hartnoll:2008rs}. For type IIB the metric is now given by
\bea\label{iibsolgen}
ds^2&=&\frac{dr^2}{r^2}+r^2\left[2dx^+dx^-+dx_1^2+dx_2^2\right]+ds^2(SE_5)+r^2\left[2C
dx^+ +h (dx^+)^2\right]
\nn
%F_5&=&2r^3dx^+\wedge dr\wedge dx^-\wedge dx_1\wedge dx_2+Vol(SE_5)\nn
%&-&\frac{1}{4}dx^+\wedge\left[*_{CY_3} dC+d(r^4C)\wedge dx_1\wedge dx_2\right]
\eea with the five-form unchanged from \reef{iibsol}. The conditions
on the one-form $C$ are as before and we demand that $h$ is a harmonic
function
on the $CY_3$ cone:
\be \nabla^2_{CY} h=0\ . \ee Choosing $h$ to have weight
$\lambda'$ under $r\partial_r$ we take \be\label{exph}
h=r^{\lambda'} f'\ , \ee where $f'$ is an eigenfunction of the
Laplacian on $SE_5$ with eigenvalue $k'$ \be -\nabla^2_{SE_5}f'=k'f'
\ee with $k'=\lambda'(\lambda'+4)$. If we set $C=0$ and
choose the branch $\lambda'=-2+\sqrt{4+k'}$ then these are the
solutions constructed in section 5 of \cite{Hartnoll:2008rs} and
have dynamical exponent $z=\tfrac{1}{2}\sqrt{4+k'}$. As noted in
\cite{Hartnoll:2008rs} an application of Lichnerowicz's theorem
\cite{lich},\cite{obata} implies that these solutions have $z\ge
3/2$ with $z=3/2$ only possible for $S^5$. Now if there
is a scalar eigenfunction with eigenvalue $k'$ and a one-form
eigenmode of the Laplacian on $SE_5$ with eigenvalue $\mu$ that
satisfy $z=\tfrac{1}{2}\sqrt{4+k'}=1+\sqrt{1+\mu}$ then we can superpose
the solution with $h$ as in \reef{exph} and the one-form $C$ as in
\reef{newcee} and have a solution with scaling symmetry with this
value of $z$. For example on $S^5$,
using the notation as before, we have $k'=l'(l'+4)$, $l'=1,2,\dots$ and
$\mu=(s+1)(s+3)$, $s=1,2,\dots$ and hence we must demand
that $l'=2(s+2)$, $s=1,2,\dots$, giving solutions with $z=3+s$.

The story for $D=11$ is very similar. The metric is now given by
\bea
ds^2&=&\frac{dr^2}{r^2}+r^4\left[2dx^+dx^-
+dx^2\right]+ds^2(SE_7)+r^4 \left[2C dx^+ +h (dx^+)^2\right]
\eea
with the four-form unchanged from \reef{11sol}. The conditions
on the one-form $C$ are as before and we demand that $h$ is a harmonic
function on the $CY_4$ cone:
\be \nabla^2_{CY} h=0\ . \ee Choosing $h$ to have
weight $\lambda'$ under $r\partial_r$ we take
\be\label{exph11} h=r^{\lambda'} f'\ , \ee where $f'$ is an
eigenfunction of the Laplacian on $SE_7$ with eigenvalue $k'$ \be
-\nabla^2_{SE_7}f'=k'f' \ee with $k'=\lambda'(\lambda'+6)$. If we
set $C=0$ and chose the branch $\lambda'=-3+\sqrt{9+k'}$
then these solutions have dynamical exponent
$z=\tfrac{1}{4}(1+\sqrt{9+k'})$. Lichnerowicz's theorem
\cite{lich},\cite{obata} implies that these solutions have $z\ge
5/4$ with $z=5/4$ only possible for $S^7$. If there is
a scalar eigenfunction with eigenvalue $k'$ and a one-form eignemode
of the Laplacian on $SE_7$ with eigenvalue $\mu$ that satisfy
$z=\tfrac{1}{4}(1+\sqrt{9+k'})=1+\tfrac{1}{2}\sqrt{4+\mu}$ then
we can superpose the solution with $h$ as in \reef{exph11} and the
one-form $C$ as in \reef{newcee11} and have a solution with scaling
symmetry with this value of $z$. For example on $S^7$, using the
notation as before,
we have $k'=l'(l'+6)$, $l'=1,2,\dots$ and
$\mu=s(s+6)+5$, $s=1,2,\dots$ and hence we must demand
that $l'=2(s+3)$, $s=1,2,\dots$,
giving solutions with $z=\tfrac{1}{2}(5+s)$.

\subsection*{Acknowledgements}
We would like to thank Seok Kim, James Sparks, Oscar Varela and Daniel Waldram.
for helpful discussions. JPG is supported by an EPSRC Senior Fellowship and a Royal Society Wolfson Award.

\end{document}